\documentclass[sn-basic]{sn-jnl}

\usepackage{graphicx}%
\usepackage{multirow}%
\usepackage{amsmath,amssymb,amsfonts}%
\usepackage{amsthm}%
\usepackage{mathrsfs}%
\usepackage[title]{appendix}%
\usepackage{xcolor}%
\usepackage{textcomp}%
\usepackage{manyfoot}%
\usepackage{booktabs}%
\usepackage{algorithm}%
\usepackage{algorithmicx}%
\usepackage{algpseudocode}%
\usepackage{listings}%

\theoremstyle{thmstyleone}%
\theoremstyle{thmstyletwo}%

\theoremstyle{thmstylethree}%

\begin{document}

\title[Dynamical evolution and rheological behavior of a sheared compound
 vesicle with a highly viscous core: a mesoscale particle-based study]{Dynamical evolution and rheological behavior of a sheared compound
 vesicle with a highly viscous core: a mesoscale particle-based study}


\author*[1]{\fnm{Antonio} \sur{Lamura}}\email{antonio.lamura@cnr.it}

\affil*[1]{\orgdiv{Istituto per le Applicazioni del Calcolo}, \orgname{CNR}, \orgaddress{\street{Via Amendola 122/D}, \city{Bari}, \postcode{70126}, \country{Italy}}}


\abstract{We study numerically the dynamics
and the rheology of a two-dimensional compound vesicle in wall-bounded
shear flow, as a minimal model of a nucleated cell. We consider the configuration in which
the annular fluid is iso-viscous with the external
one, while the inner core is more viscous. 
Increasing the core
viscosity drives the external vesicle away from steady tank-treading at values $\phi \geq 0.6$. When
an internal body is present, the annular fluid partially shields the
outer membrane from the viscous core, so that at small filling fractions the external vesicle
retains tank-treading with a higher inclination angle and the intrinsic
viscosity stays comparatively low. The internal vesicle, by contrast,
tumbles and migrates away from the center, and at high $\phi$ this
motion couples back to the host, producing intermittent shape
excursions of the external membrane. The inner viscous body thus acts
as an internal governor of the dynamical state.}




\maketitle

\section{\label{sec:intro}Introduction}

Eukaryotic cells are structurally complex systems. They are typical examples of
nucleated cells containing a nucleus that is considerably more viscous than the surrounding cytoplasm. The nucleus is
enclosed by its own membrane and embedded within the plasma membrane,
resulting in a compound, two-compartment system bounded by two distinct
interfaces. This internal architecture plays a central role in cell mechanics
and motility. As the largest and mechanically dominant organelle, the nucleus
can represent the rate-limiting element when cells deform through
microvascular constrictions, migrate through interstitial tissues, or undergo
mechanical characterization in deformability-cytometry devices
\citep{linc04,davi15}.
One prominent example of eukaryotic cells is provided by leukocytes (white
blood cells - WBC) which play a relevant role in the immune system during
inflammatory response. In some conditions, the concentration of WBC becomes
very high as in the case of leukemia or in cancer tissues. Under these
circumstances, WBC deeply affect blood flow in large vessels \citep{schm80}.
It is thus relevant to understand their dynamics and rheological properties
which have not been fully understood due to their complex internal structure.
A minimal physical model that captures these essential features - namely,
a more viscous inner body enclosed by a membrane and suspended within an
outer membrane - therefore provides a useful framework for shedding light on
the mechanical contribution of the nucleus.

Lipid-bilayer vesicles constitute a minimal model for the outer
cellular boundary. Their dynamics are governed by membrane bending rigidity,
fluidity and local inextensibility, together with a fixed enclosed area.
Even a single vesicle can exhibit a broad range of nonequilibrium dynamical
states under shear flow, including tank-treading, tumbling, and the
intermediate trembling regime. The selected dynamical
mode depends primarily on the viscosity ratio between the inner fluid and 
the suspending one, as well as on the reduced area
\citep{kell82,seif97,nogu04,kant05,kant06,misb06,nogu07bis,fink08,mess09,
vlah09,abre14,lamu22}.
Compound vesicles, containing an
additional enclosed body, introduce a further level of mechanical complexity.
The inner phase occupies a finite fraction of the host volume and possesses
its own viscosity, thereby modifying the dynamics of the outer object.
The internal body may migrate away from the center, develop an independent
dynamical state, and strongly influence the overall response of the compound
system. As a result, the relevant parameter space is substantially expanded
\citep{veer11,kaou13,leva14,sinh19,kuma24,lamu25}.

A suitable description of this system requires a hydrodynamic framework
that incorporates thermal fluctuations, naturally accommodates impenetrable
boundaries, and, most importantly, permits the viscosity of each of the
three fluid compartments separated by the two membranes to be
controlled independently. Multi-particle collision (MPC) dynamics satisfies
these requirements. It is a mesoscale, momentum-conserving particle-based
method whose transport properties can be tuned through the collision rule
and which couples naturally to molecular-dynamics descriptions of deformable
membranes \citep{male99,male00,kapr08,gomp09}.
Appropriate boundary conditions at solid walls and
internal interfaces can be conveniently implemented
\citep{lamu01,lamu02,mess09}.
In addition, an angular-momentum-conserving collision rule can be
employed to represent
torques acting on suspended objects and thereby capture the balance
between tank-treading and tumbling dynamics more faithfully
\citep{nogu07,goetze07,lamu13,afik16,lamu22}.
In the present model, each membrane is represented as a closed chain of
beads interacting through bonding and curvature potentials, while the
three fluid compartments are described by MPC particle populations with
independently prescribed viscosities \citep{lamu25}.
Here, we investigate the biologically relevant configuration in which the
annular fluid is iso-viscous with the external one, while the enclosed
core is more viscous. This arrangement provides a minimal
two-dimensional representation of a nucleated cell in which the nucleus is
more viscous than the surrounding cytoplasm. We examine how the increased
core viscosity influences the dynamic response and the rheological behavior
in wall-bounded shear flow under different conditions.

Previous studies \citep{veer11,sinh19} were limited to consider a solid-like inner core 
in a quasi-spherical vesicle under shear flow. A transition was observed of the
external vesicle from tank-treading to tumbling motion by increasing the
size of the spherical internal core. A non-spherical inclusion was capable
of inducing a more complex dynamics during which the vesicle performed
a swinging motion. The amplitude of this modulated motion was found to increase
with a larger inclusion size \citep{veer11}.
In the present investigation, non-circular vesicles are considered by varying
their relative degree of swelling. The viscosity of the inner vesicle is changed
over a wide range while still keeping it large but finite.
Moreover, high filling fractions of the internal vesicle
are considered. This allows us to observe a rich and complex dynamical
behavior which is fully characterized by investigating the inclination angle
and the asphericity of the vesicles. The rheological behavior is also considered and
related to the observed dynamics.

\section{Model and simulation method}
\label{sec:model}

\subsection{Membrane model}

The compound vesicle is modeled in two dimensions by an external and an internal
membrane. Each membrane is made of a closed chain of beads, whose spatial position
is denoted by $\mathbf{r}_{i}$,
of diameter $r_v$ and mass $M$. Neighboring beads interact
through an elastic potential
\begin{equation}
  U_{\rm bond} = \frac{k_h}{2 r_0^2}\sum_{i}\left(|\mathbf{r}_{i+1}-\mathbf{r}_i|-r_0\right)^2 ,
  \label{eq:bond}
\end{equation}
with bond stiffness $k_h$ and average bond length $r_0$,
which with a large stretching modulus enforces the local
inextensibility of the membrane. The resistance to bending is provided
by a curvature potential
\begin{equation}
  U_{\rm bend} = \frac{\kappa}{r_0} \sum_{i}\bigl(1-\cos\theta_i\bigr),
  \label{eq:bend}
\end{equation}
where $\theta_i$ is the angle between consecutive bond vectors and
$\kappa$ is the bending rigidity.
The area enclosed by each membrane is kept fixed at its target value by
a harmonic area constraint
\begin{equation}
  U_{\rm area} = \frac{k_A}{2 r_0^4}\,(A-A_0)^2,
  \label{eq:area}
\end{equation}
with $A$ the instantaneous enclosed area, $A_0$ its target, and $k_A$ a
stiff area-compression modulus.
Excluded-volume interactions, modeled by a shifted Lennard-Jones potential
\begin{equation}
  U_{\rm rep} = 4 \epsilon \left [ \left(\frac{\sigma}{r}\right)^{12}-\left(\frac{\sigma}{r}\right)^6\right ] + \epsilon
  \label{eq:wca}
\end{equation}
truncated at its minimum $r_{cut}$, prevent the membranes from crossing each other.
Membranes are simulated by molecular dynamics
with a velocity-Verlet integrator of time step
$\Delta t_{MD}$~\citep{lamu22,lamu25}.

The degree of swelling of each vesicle is quantified by its reduced
area
\begin{equation}
  A^{*} = \frac{A_0}{\pi R_0^{2}},
  \qquad R_0 = \frac{L_0}{2\pi},
  \label{eq:reducedarea}
\end{equation}
where $L_0$ is the membrane contour length, so that $A^{*}=1$ corresponds
to a circle and smaller values to increasingly deflated shapes. We
consider an external membrane with $A^{*}_{ext}=0.7$ or $0.8$ and an
internal one with $A^{*}_{int}=0.8$. The filling fraction of the
compound vesicle is defined as \citep{kaou13,leva14,sinh19}
\begin{equation}
  \phi = \frac{R_{0,int}}{R_{0,ext}} .
  \label{eq:phi}
\end{equation}
The values $\phi = 0,\,0.2,\,0.4,\,0.6,\,0.8$ will be considered in the following. 
The limiting case $\phi=0$ contains no internal vesicle and reduces to a simple
vesicle whose entire interior has viscosity contrast $\lambda_{in}$; it
serves throughout as the reference against which the effect of the inner
body is measured.

\subsection{Solvent: multi-particle collision dynamics}

The three fluid compartments are described by multi-particle collision
dynamics ~\citep{male99,kapr08,gomp09}, a
mesoscale, momentum-conserving particle method that reproduces
hydrodynamics with thermal fluctuations. The solvent is represented by
point-like particles of mass $m$ that evolve in alternating streaming and
collision steps. In the streaming step, particles move ballistically for
a time $\Delta t$; in the collision step, particles are sorted into a
grid of square cells of size $a$ and their relative velocities are
rotated about the cell center of mass. We employ the
angular-momentum-conserving variant of the collision
rule~\citep{goetze07,nogu07} so that torques on the suspended
membranes are captured correctly while keeping temperature
constant \citep{alla02}.

The key feature of the model is that the viscosity $\eta$ of each of the three
compartments - outer (out) fluid, annular (an) fluid, and inner (in) core - is
prescribed independently. Particles are tagged according to the
compartment in which they reside (delimited by the two membranes), and
the collision rule is applied compartment-wise so that each region
carries its own transport coefficients. In particular, viscosities are changed
independently by varying solvent mass (see Appendix~\ref{app:visc}).
The annular fluid is taken
iso-viscous with the external one, $\lambda_{an}=\eta_{an}/\eta_{out}=1$, while the inner
core is more viscous, with contrast
$\lambda_{in} = \eta_{in}/\eta_{out}$
varied in the range $[1,15]$, $\eta_{out, an, in}$ being the viscosities in the three
compartments.

Membrane beads are coupled to the solvent by 
three-body collisions in order to ensure conservation of linear and angular momenta
\citep{mess09} (full details can be found elsewhere \citep{lamu22}).

\subsection{Geometry and parameter setting}

The vesicle is confined between two parallel walls separated by a gap
$L_y=5.76 R_{0,ext}$ in the velocity-gradient ($y$) direction, with
periodic boundary conditions along the flow ($x$) direction over a
length $L_x=12.04 R_{0,ext}$, being $R_{0,ext}=15.28 a$. Planar Couette flow of shear rate
$\dot\gamma$ is imposed by moving the two walls in opposite directions
with speed $\pm \dot\gamma L_y/2$; no-slip is enforced at the walls by
bounce-back \citep{lamu01,lamu02}. The wall velocity is chosen so as to ensure that inertial effects
are negligible by keeping the Reynolds number smaller than $0.1$. 
Moreover, compressibility effects are maintained under control by limiting the upper value of the Mach number to $0.25$ \citep{lamu02}. The reduced shear rate is set to
$\dot\gamma^*= \dot\gamma \tau_R=20$ where $\tau_R=\eta_{out} R_{0,ext}^3/\kappa$ is the
relaxation time of the external membrane.

We choose $M=3 m_{out}$, $\Delta t / \Delta t_{MD}=64$, $r_v=r_0=a/10$,
$r_{cut}=a$, $k_h=3 \times 10^2 k_B T$, $\kappa_{ext}=\kappa_{int}=3.27 k_B T R_{0,ext}$, 
$k_A=0.4 k_B T$, and $\epsilon=10 k_B T$,
where $k_B T$ is the thermal energy.

\subsection{Observables}
\label{sec:observables}

The orientation and shape of each vesicle are extracted from the
gyration tensor of its membrane beads,
\begin{equation}
  G_{\alpha\beta}=\frac{1}{N}\sum_{i}
  (r_{i,\alpha}-r_{cm,\alpha})(r_{i,\beta}-r_{cm,\beta}),
  \qquad \alpha,\beta\in\{x,y\},
  \label{eq:gyr}
\end{equation}
where $\mathbf{r}_{cm}$ is the position of the vesicle center of mass,
with eigenvalues $\xi_1\ge\xi_2$ and corresponding
eigenvectors. The inclination angle $\theta$ is the angle between the
eigenvector of the largest eigenvalue $\xi_1$ and the flow
direction, measured in the interval $[-\pi/2,\pi/2]$ and normalized as
$\theta/\pi$. The asphericity
\begin{equation}
  A=\left(\frac{\xi_1-\xi_2}{\xi_1+\xi_2}\right)^{2}
  \label{eq:aspher}
\end{equation}
vanishes for a circle and increases towards unity for elongated shapes.
The tank-treading frequency $\omega$ is obtained from the rotation of
tagged membrane beads around the vesicle center of mass
and reported as $\langle\omega\rangle/\dot\gamma$. Finally, the rheological
response is characterized by the intrinsic viscosity
\begin{equation}
  \langle\eta_I\rangle=\frac{\langle\eta_{\rm eff}\rangle-\eta_{out}}{\eta_{out}\,c},
  \label{eq:intrinsic}
\end{equation}
where $\eta_{\rm eff}$ is the effective system viscosity evaluated as
$\eta_{\rm eff}=\sigma_{xy}/\dot\gamma$ where $\sigma_{xy}$ is the $xy$ component of the stress tensor at walls \citep{mewi11},
and $c=A_{0,ext}/(L_x L_y)$ is the external vesicle area fraction. 
Angular brackets
denote time averages in the steady state.

\section{Results and discussion}
\label{sec:results}

We first characterize the dynamical state of the external
vesicle through its mean inclination angle and tank-treading frequency
(Sec.~\ref{sec:states}), then turn to the rheology
(Sec.~\ref{sec:rheo}), and finally analyze the coupled internal--external
dynamics, the migration of the inner body, and the resulting shapes
in the case of the more viscous inner vesicle (Sec.~\ref{sec:coupling}). Two families of geometries are compared
throughout: a deflated host, $A^{*}_{ext}=0.7$ with $A^{*}_{int}=0.8$,
and a rounder host, $A^{*}_{ext}=A^{*}_{int}=0.8$, while varying the filling fraction.

\subsection{Dynamical states of the external vesicle}
\label{sec:states}

\begin{figure}[ht]
\begin{center}
\includegraphics*[width=0.65\textwidth,angle=0]{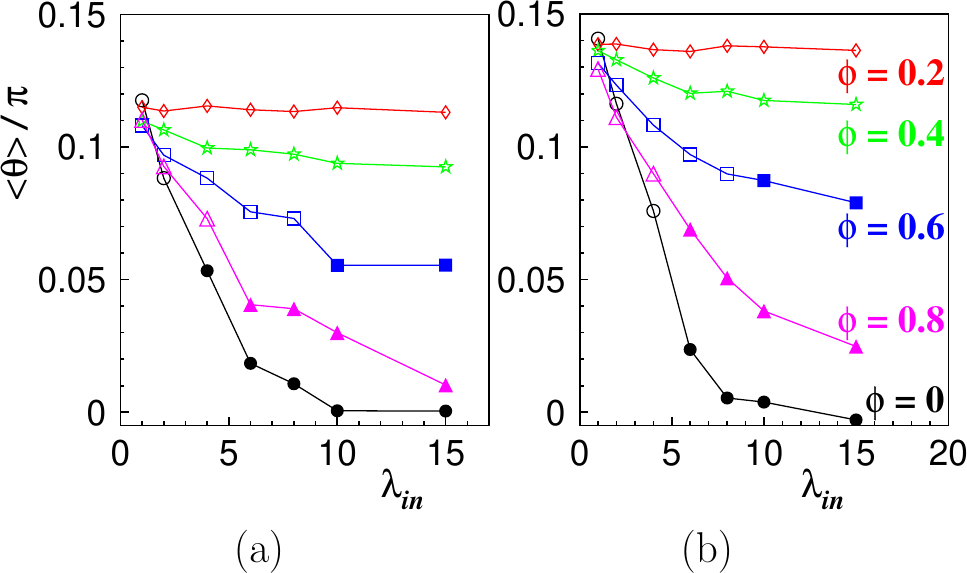}
\end{center}
\caption{\label{fig1} Dependence of the average inclination angle of the
external vesicle on $\lambda_{in}$
for different values of the filling fraction $\phi$ 
in the cases with (a) $A^*_{ext}=0.7, A^*_{int}=0.8$ and (b) $A^*_{ext}=A^*_{int}=0.8$.
Empty symbols correspond to states with tank-treading motion.}
\end{figure}

Figure \ref{fig1} shows the mean inclination angle
$\langle\theta\rangle/\pi$ of the external vesicle as a function of the
core viscosity contrast $\lambda_{in}$ for the two geometries. For the
reference simple vesicle ($\phi=0$) the inclination angle decreases
steeply with $\lambda_{in}$ and approaches zero: as the interior becomes
more viscous the vesicle leaves the steady tank-treading regime and progressively
begins to tumble, in agreement with the classical behavior of viscous
vesicles in shear flow~\citep{kant06,misb06,nogu07bis,mess09}. This
transition is signaled in the figure by the change from open symbols
(tank-treading) to filled ones. Introducing an internal body
suppresses this decay: the smaller the filling fraction
$\phi$, the higher the inclination angle remains, and at $\phi=0.2, 0.4$ 
the external vesicle keeps a large, only weakly
$\lambda_{in}$-dependent inclination angle over the whole range of core
viscosities persisting in tank-treading. The annular fluid is iso-viscous with the exterior and
screens the outer membrane from the highly viscous core. As a consequence the
outer object behaves as a low-viscosity vesicle even when the nucleus is
stiff but small in size. 
When increasing the size of the inner core, the inclination of the external vesicle
diminishes at fixed viscosity contrast \citep{kaou13,sinh19} and the transition to a non-tank-treading state is observed, when  $\phi \geq 0.6$, at values of 
$\lambda_{in}$ which reduce with $\phi$. 
The same dependence with $\phi$ is found for both host geometries;
the rounder host ($A^{*}_{ext}=0.8$, panel~b) presents a slightly larger
steady orientation with respect to the more deflated one
($A^{*}_{ext}=0.7$, panel~a), consistent with the general observation
that elongated vesicles tank-tread at a smaller inclination
than quasi-circular ones. 

\begin{figure}[ht]
\begin{center}
\includegraphics*[width=0.65\textwidth,angle=0]{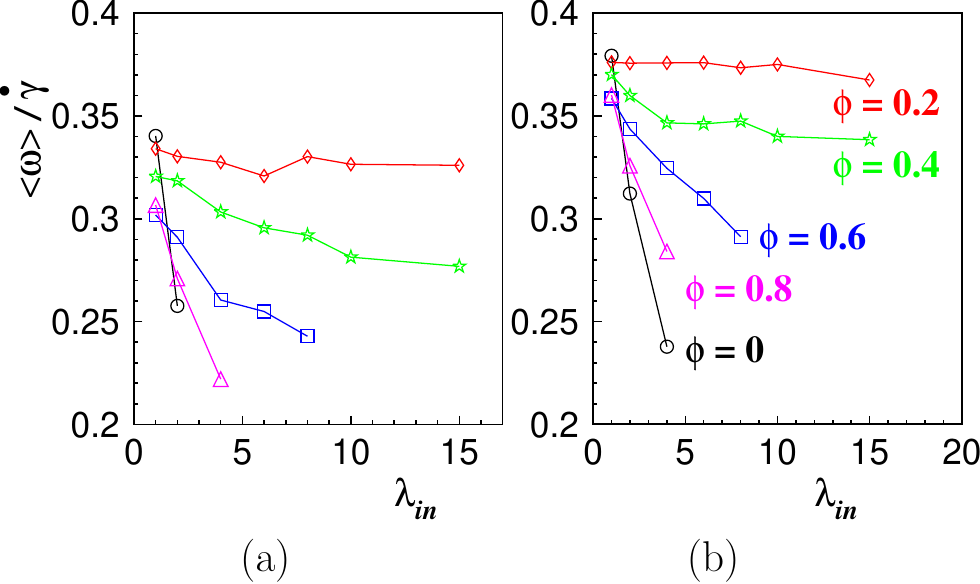}
\end{center}
\caption{\label{fig2} Dependence of the average tank-treading frequency
of the external vesicle on $\lambda_{in}$
for different values of the filling fraction $\phi$ 
in the cases with (a) $A^*_{ext}=0.7, A^*_{int}=0.8$ and (b) $A^*_{ext}=A^*_{int}=0.8$.}
\end{figure}

The tank-treading frequency, Fig.~\ref{fig2}, is consistent with what was already observed. 
The value of $\langle\omega\rangle/\dot\gamma$ decreases with increasing
$\lambda_{in}$ for every $\phi$ since the more viscous interior is
harder to circulate. Here the decrease is extremely slow for small sizes and
strongest for large
filling fractions. At low $\phi$, the
membrane keeps tank-treading at a comparatively high frequency with respect to the
single vesicle case since the presence of the inner core is screened by the large
amount of the annular area. As the inner size increases, the frequencies
reduce approaching the values of the case with $\phi=0$, as observed in the case
with a solid inclusion \citep{veer11}.
Frequencies display larger values for more rounded host vesicles. 

\subsection{Rheology}
\label{sec:rheo}

\begin{figure}[ht]
\begin{center}
\includegraphics*[width=0.65\textwidth,angle=0]{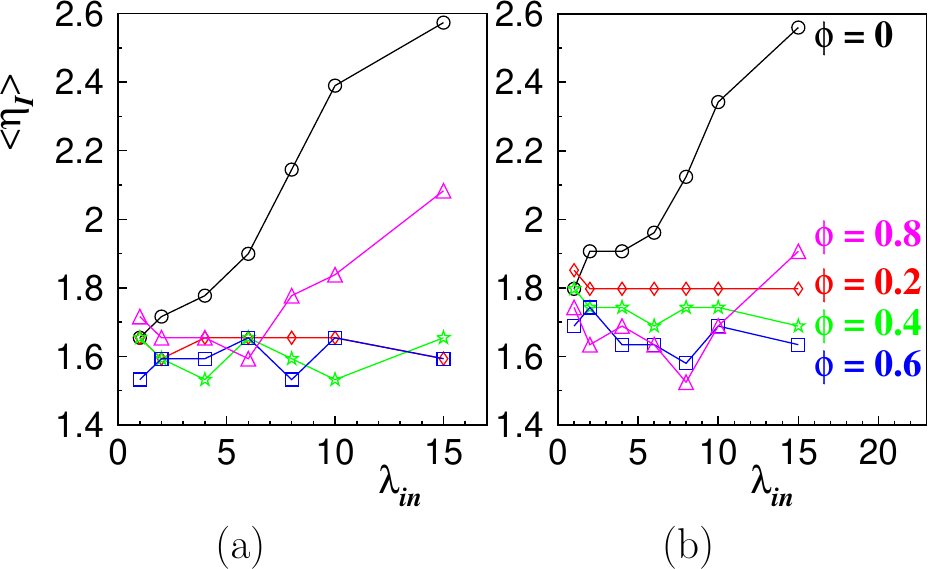}
\end{center}
\caption{\label{fig3} Dependence of the average intrinsic viscosity
on $\lambda_{in}$
for different values of the filling fraction $\phi$ 
in the cases with (a) $A^*_{ext}=0.7, A^*_{int}=0.8$ and (b) $A^*_{ext}=A^*_{int}=0.8$.}
\end{figure}

The intrinsic viscosity, shown in Fig.~\ref{fig3}, is related to the
observed dynamical states of Sec.~\ref{sec:states}. For the simple vesicle ($\phi=0$) 
$\langle \eta_{I} \rangle$ increases
with $\lambda_{in}$ due to the loss of tank-treading state and the 
onset of tumbling, together with the increasing internal
dissipation \citep{lamu22}. 

When an inner body is present, the intrinsic viscosity remains
lower than the $\phi=0$ values at the same $\lambda_{in}$ over the whole explored range,
and  for $\phi \leq 0.6$ it varies only weakly with $\lambda_{in}$.
At these filling fractions the external membrane keeps a tank-treading or swinging motion 
(see next Section)
being separated 
from the viscous core by the iso-viscous annulus. 
The differences among the curves for $\phi \leq 0.6$ are comparable to the scatter of data within error 
uncertainties,
so it is difficult to argue a dependence on $\phi$ in this range.
 The only case showing a clear increase of the intrinsic viscosity is
$\phi=0.8$ where $\langle \eta_{I} \rangle$ rises at large
$\lambda_{in}$.
As shown in the next Section, for $\phi=0.8$ and large $\lambda_{in}$ both vesicles undergo intermittent 
tumbling and morphology fluctuations of the host. This is consistent with an increase of dissipation due
to strong shape changes, as previously pointed out for vesicle suspensions \citep{kant08}.

Overall, the presence of a viscous inner body reduces the intrinsic viscosity with respect to a simple vesicle
with the same interior viscosity contrast. This reduction is observed as long as the host maintains 
a steady orientation. Only when the two membranes are tightly packed and the host is driven
into shape changes, $\langle \eta_{I} \rangle$ starts to increase again with $\lambda_{in}$.

\subsection{Coupled internal--external dynamics, migration, and shapes}
\label{sec:coupling}

\begin{figure*}[ht]
\includegraphics*[width=0.99\textwidth,angle=0]{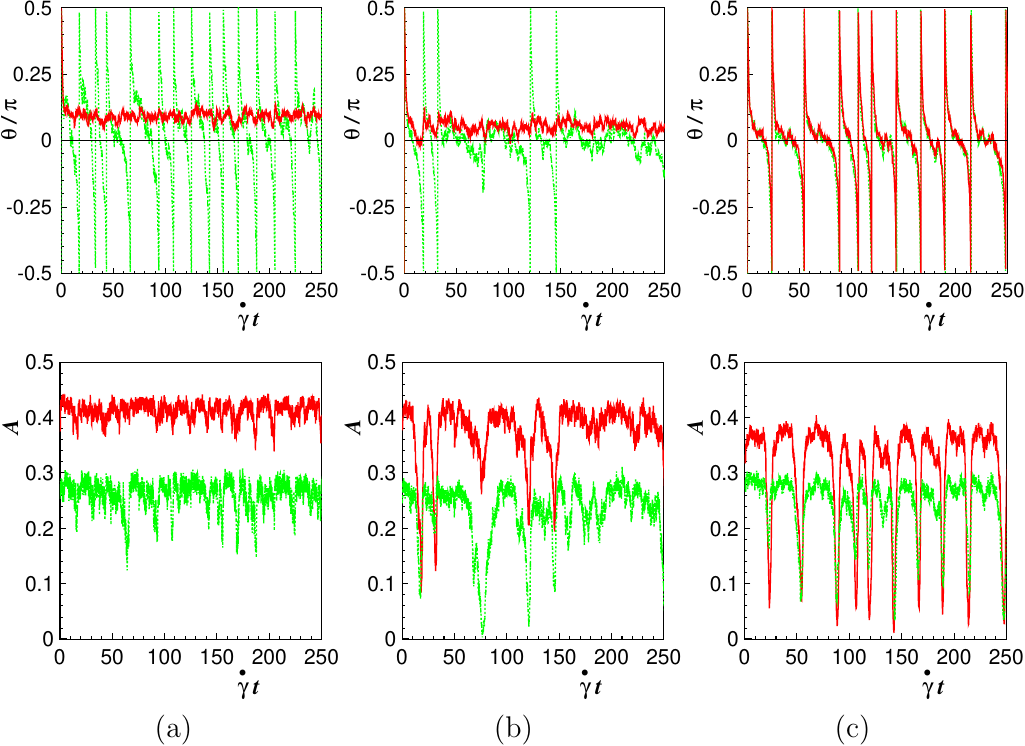}
\caption{\label{fig4} Time evolution of (upper row) the inclination angles
$\theta$ and (lower row) the asphericities $A$ for the external (red continuous line)
and internal (green dashed line) vesicles at filling fractions $\phi=0.4$ (a), $0.6$ (b), $0.8$ (c) with $\lambda_{in}=15$ and $A^*_{ext}=0.7, A^*_{int}=0.8$.
}
\end{figure*}

The averaged quantities above hide a rich time-dependent behavior that
is revealed by the instantaneous inclination angles and asphericities of
the two membranes. Figure~\ref{fig4} shows these
time series at the largest core viscosity $\lambda_{in}=15$ for filling
fractions $\phi=0.4,\,0.6,\,0.8$ in the case of the deflated host
($A^{*}_{ext}=0.7$).
\begin{figure}[ht]
\begin{center}
\includegraphics*[width=0.65\textwidth,angle=0]{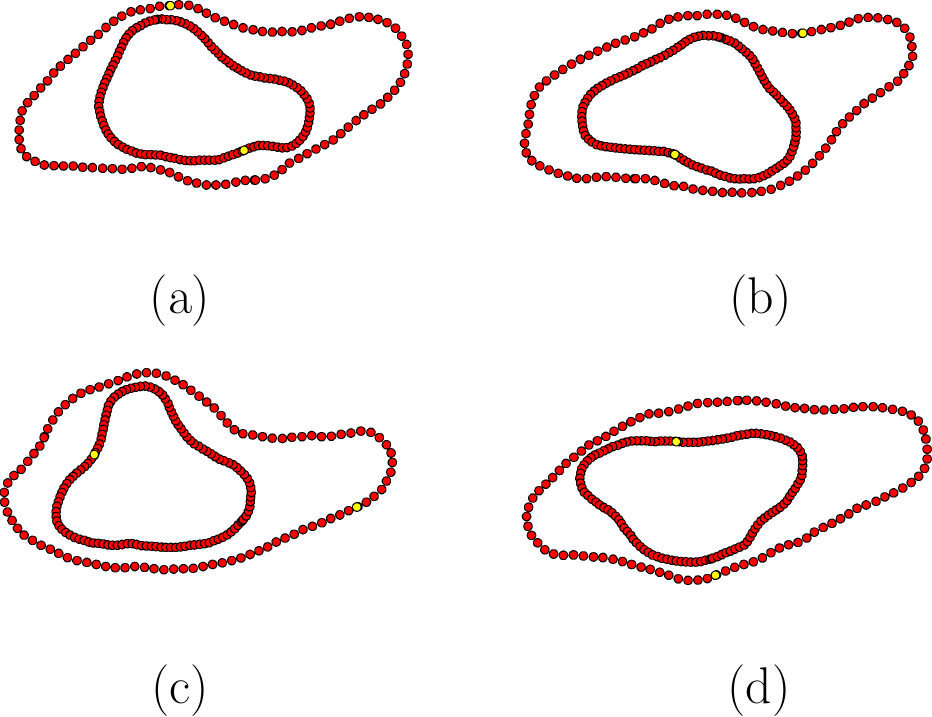}
\end{center}
\caption{\label{fig5} Configurations of compound vesicles, corresponding to the
case of Fig.~\ref{fig4} (b), for filling fraction $\phi=0.6$
with $\lambda_{in}=15$ and $A^*_{ext}=0.7, A^*_{int}=0.8$. Snapshots are taken at times
$\dot\gamma t= 67.5$ (a), $72.5$ (b), $77.5$ (c), $82.5$ (d).}
\end{figure}
\begin{figure}[ht]
\begin{center}
\includegraphics*[width=0.6\textwidth,angle=0]{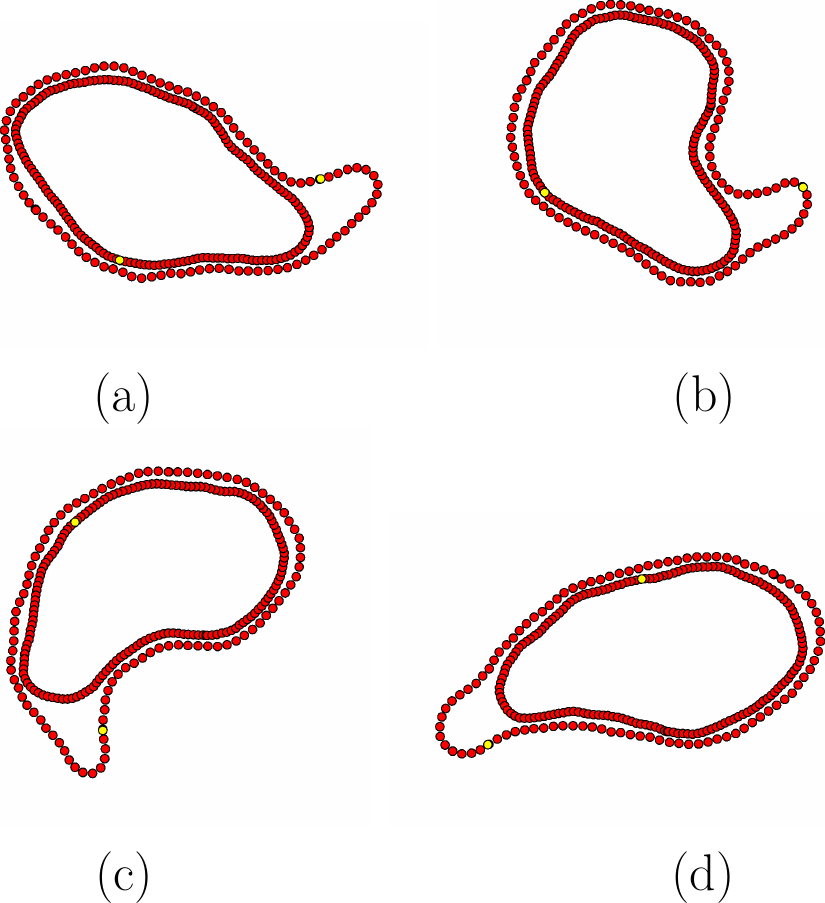}
\end{center}
\caption{\label{fig6} Configurations of compound vesicles, corresponding to the
case of Fig.~\ref{fig4} (c), for filling fraction $\phi=0.8$
with $\lambda_{in}=15$ and $A^*_{ext}=0.7, A^*_{int}=0.8$. Snapshots are taken at times
$\dot\gamma t= 85$ (a), $87.5$ (b), $90$ (c), $92.5$ (d).}
\end{figure}
\begin{figure*}[ht]
\includegraphics*[width=0.99\textwidth,angle=0]{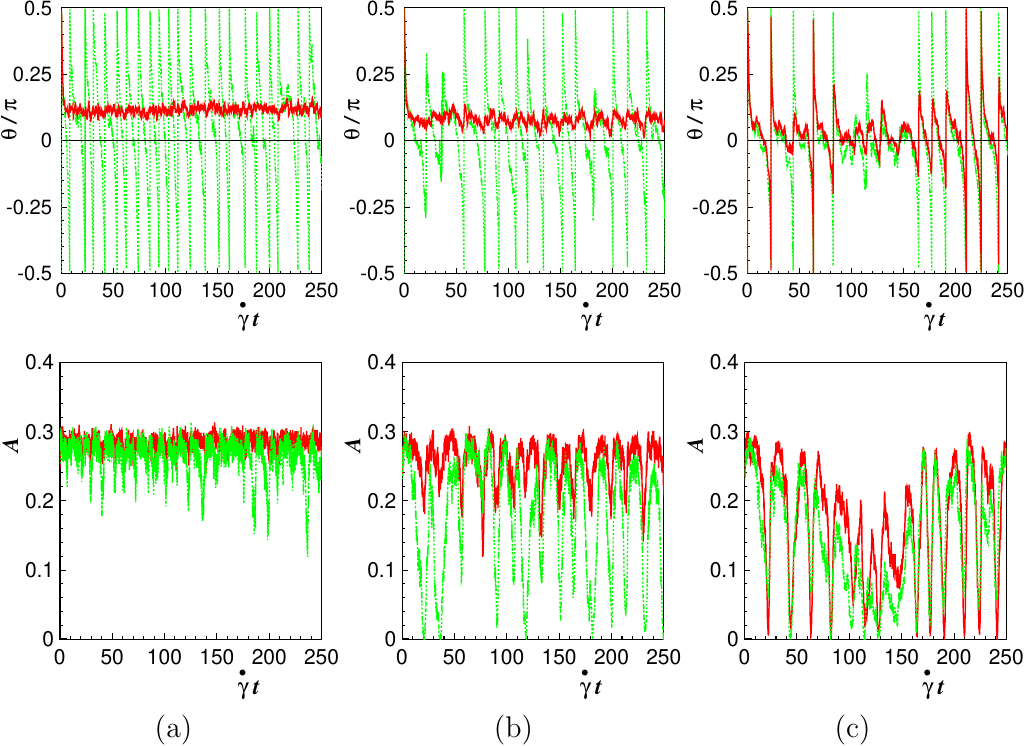}
\caption{\label{fig7} Time evolution of (upper row) the inclination angles
$\theta$ and (lower row) the asphericities $A$ for the external (red continuous line)
and internal (green dashed line) vesicles at filling fractions $\phi=0.4$ (a), $0.6$ (b), $0.8$ (c) with $\lambda_{in}=15$ and $A^*_{ext}=A^*_{int}=0.8$.}
\end{figure*}
\begin{figure}[ht]
\begin{center}
\includegraphics*[width=0.65\textwidth,angle=0]{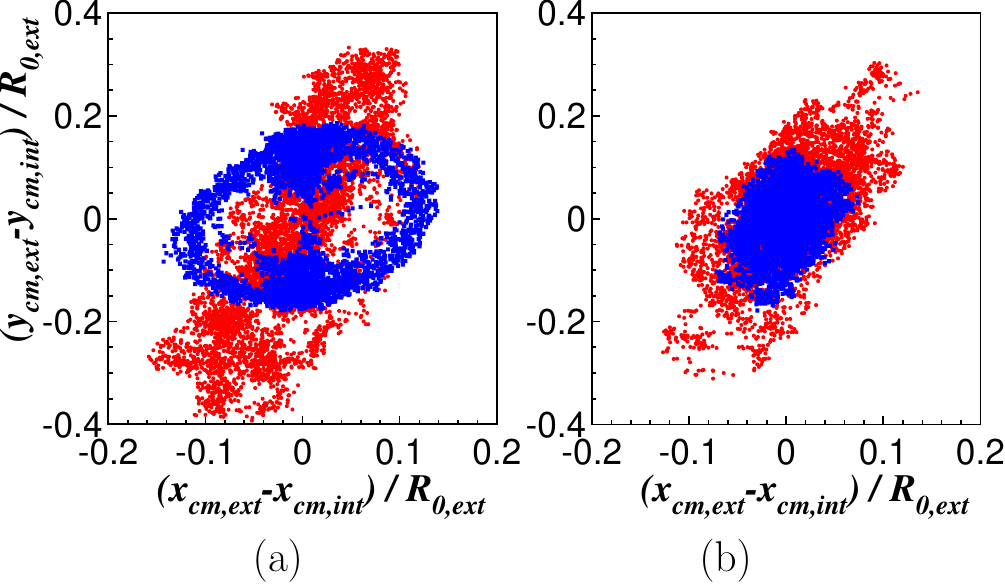}
\end{center}
\caption{\label{fig8} Probability distribution of the center-of-mass relative positions
for a compound vesicle
corresponding in panels (a) and (b) to the cases shown  
in Fig.~\ref{fig4} (b)-(c)
and in Fig.~\ref{fig7} (b)-(c), respectively. Red and blue symbols correspond to the filling
fractions $\phi=0.6, 0.8$, respectively.}
\end{figure}

At the lowest filling fraction shown, $\phi=0.4$---Fig.~\ref{fig4}(a), the two membranes are
essentially decoupled: the external vesicle (red) tank-treads at a
steady positive inclination angle with an almost constant asphericity,
while the internal vesicle (green), being small and highly viscous,
tumbles---its inclination angle sweeps repeatedly through the full range
and its asphericity oscillates. As $\phi$ increases to $0.6$---Fig.~\ref{fig4}(b),
the inner body occupies a larger fraction of the host and the two
motions become coupled: the external vesicle shows now swinging motion with its
main axis oscillating around a positive mean angle \citep{leva14}. 
The inner vesicle exhibits trembling motion
during which the main axis oscillates around the flow direction with strong shape changes
\citep{kant07} while some tumbling events rarely occur.
The external inclination angle develops
local minima when the internal tilt assumes negative values
and its asphericity shows pronounced dips that
are synchronized with the shape fluctuations of the internal vesicle. 
Representative configurations
in the time range $67.5 \leq \dot\gamma t \leq 82.5$
are shown in Fig.~\ref{fig5}. The snapshots
make explicit how the dynamics of the off-center inner body deforms the host:
the outer contour bulges and relaxes as the internal
vesicle buckles assuming a trilobate shape. 
At $\phi=0.8$---Fig.~\ref{fig4}(c), both vesicles undergo synchronized tumbling motions. During
the reorientation when vesicles acquire positive inclinations, their shapes become more rounded as
both asphericities show local minima. 
Figure \ref{fig6} illustrates, in the time range $85 \leq \dot\gamma t \leq 92.5$,
how the two membranes are in
close contact over much of the perimeter and the host is driven into a less
elongated and more compact shape during the reorientation of the inner vesicle while
keeping a protrusion where most of the annular fluid is accommodated.

The time dependencies of orientation and asphericity in the case of the rounder host 
are shown in Fig.~\ref{fig7}.
For the smaller inclusion with $\phi=0.4$---Fig.~\ref{fig7}(a), the qualitative
behavior does not change with respect to the case of the more deflated external vesicle:
the outer membrane tank-treads while the inner one tumbles. By increasing the size, it is
possible to observe at $\phi=0.6$---Fig.~\ref{fig7}(b), that the internal vesicle exhibits
full tumbling while the external one swings. During its flipping, it deforms gaining a rounded shape while forcing an in-phase
deformation of the external vesicle, as it appears in the lower panel Fig.~\ref{fig7}(b).
The typical evolution in one of these cycles is clearly shown in the supplementary
animation (Online Resource 1).
At the largest filling fraction, $\phi=0.8$---Fig.~\ref{fig7}(c),
both membranes undergo large, nearly in-phase variations of the inclination
showing an alternation of tumbling and trembling motion.
The external asphericity drops periodically towards small values, indicating
that the host is repeatedly deformed into a more compact shape as the
internal body switches from negative to positive orientation. This can be visualized
in the supplementary animation (Online Resource 2). 
The intermittent loss of steady orientation seen
here is consistent with the upturn of the intrinsic viscosity
at $\phi=0.8$ noted in Sec.~\ref{sec:rheo}. This can be related to shape fluctuations
which favor energy dissipation and may contribute to the increase of $\langle \eta_{I} \rangle$, as previously observed
in vesicles \citep{kant08}.

The relative position of the two vesicles confirms that the inner body
does not remain centered. Figure~\ref{fig8} shows the probability
distribution of the relative positions of the centers of mass for
$\phi=0.6$ (red) and $\phi=0.8$ (blue) and the two host geometries. At $\phi=0.6$ the inner body
explores a broad region and traces an extended distribution,
signaling that it orbits within the host as it tumbles.
At $\phi=0.8$
the tighter packing confines the separation to a smaller, more compact
distribution. It is interesting to note that in the case of the more deflated host,
the pattern exhibits closed trajectories consistently with the presence of the protrusion in the external
vesicle, as observed in Fig.~\ref{fig6}.

\section{\label{sec:concl}Conclusions}

We have studied numerically the dynamics and rheology of a two-dimensional compound vesicle in wall-bounded shear flow, using multi-particle collision dynamics coupled to two bead-spring membranes so that the viscosities of the outer fluid, the annulus and the enclosed core can be prescribed independently. Motivated by the architecture of nucleated cells, we considered the case of an annulus iso-viscous with the exterior and a core of higher viscosity, with 
viscosity contrast 
$\lambda_{in}$ up to 15, filling fractions $\phi$ up to 0.8, and two host geometries 
($A_{ext}^*=0.7, 0.8$ with $A_{int}^*=0.8$).

The presence of the inner body qualitatively changes how the host responds to a viscous interior. A simple vesicle whose whole interior carries the contrast 
$\lambda_{in}$
leaves tank-treading and progressively begins to tumble as $\lambda_{in}$ grows, its mean inclination angle falling towards zero and its tank-treading frequency dropping. When the viscous fluid is instead confined to a core surrounded by an iso-viscous annulus, the annular fluid screens the outer membrane: at $\phi=0.2, 0.4$ the host retains tank-treading, with a large and only 
weakly $\lambda_{in}$-dependent inclination angle, across the entire range of core viscosities examined. The screening weakens as the core size grows, and for $\phi \geq 0.6$ the transition to non-tank-treading dynamical states reappears at a threshold contrast that decreases with $\phi$. The rounder host tank-treads at a slightly larger inclination and higher frequency than the deflated one, consistent with the behavior of single vesicles.

The rheological response is closely related to the dynamical state. The intrinsic viscosity
$\langle \eta_I \rangle$ is lower than the single-vesicle value, and for $\phi \leq 0.6$, varies only weakly
with $\lambda_{in}$ within the scatter of data, consistent with a core that keeps tank-treading or swinging motion
and is separated from the inner core by the annulus.
The exception is the most tightly packed case, $\phi=0.8$, where $\langle \eta_I \rangle$ increases at large viscosity contrast, in
correspondence with intermittent tumbling and strong shape changes of the host.
These intermittent excursions are consistent with the upturn in the effective viscosity, in line with the known enhancement of dissipation by shape fluctuations in single vesicles. Throughout, the center of mass of the inner body traces an extended orbit within the host at $\phi=0.6$ and a tighter, more compact distribution at $\phi=0.8$, with a closed trajectory in the deflated host that reflects the protrusion seen in the configurations.

These results show that the enclosed viscous body strongly influences the dynamical state of the host: its size and viscosity 
determine whether the host tank-treads, swings or tumbles, without altering the physical features of
the outer membrane or the suspending fluid. 
The intrinsic viscosity follows this behavior, staying below the single-vesicle value until tight packing induces shape
excursions.
This suggests that, in engineered carriers, an internal cargo rather than the outer shell could influence how 
the object orients and deforms in flow, a possibility that should be tested in three dimensions and in channel geometries.

\backmatter

\bmhead{Supplementary information}

This article has two accompanying supplementary files:\\
1. Online Resource 1 - Animation of the time evolution of a compound vesicle
in the case with $A_{ext}^*=A_{int}^*=0.8$, $\lambda_{in}=15$, and $\phi=0.6$;\\
2. Online Resource 2 - Animation of the time evolution of a compound vesicle
in the case with $A_{ext}^*=A_{int}^*=0.8$, $\lambda_{in}=15$, and $\phi=0.8$.

\bmhead{Acknowledgements}

The work was performed under the auspices of GNFM-INdAM.

\section*{Declarations}

\begin{itemize}
\item Funding: No external funding received
\item Conflict of interest/Competing interests: None
\item Ethics approval and consent to participate: Not applicable
\item Consent for publication: Not applicable
\item Data availability: The datasets generated during and/or analysed during the current study are available from the corresponding author on reasonable request.
\end{itemize}

\begin{appendices}

\section{Control of the compartment viscosities}\label{secA1}

\label{app:visc}

In MPC dynamics, the fluid viscosity $\eta$ is a
known function of the mean number $n$ of solvent particles per collision cell,
their mass $m$, the collision time $\Delta t$, 
the cell size $a$, and the thermal energy $k_B T$
\citep{nogu08}:
\begin{equation}
    \eta=\frac{m}{\Delta t} \left [ \left ( \frac{l}{a} \right )^2  \left ( \frac{n^2}{n-1}-\frac{n}{2}\right ) + \frac{1}{24}\left ( n-\frac{7}{5}\right )    \right ] ,
\end{equation}
$l=\Delta t \sqrt{k_B T/m}$ being the solvent-particle mean-free path. By setting $n=10$ and $l/a \ll 0.1$, the first term in the square brackets is negligible and 
the value of $\eta$ is monotonic
in $m$. The viscosity of each compartment can then be set
independently by assigning the appropriate mass to the solvent
particles it contains, once the same time step $\Delta t$ has been chosen for all the MPC particles.

\end{appendices}


\bibliography{biblio_ves}

\end{document}